\documentclass[12pt]{iopart}
 \usepackage{graphicx}
 \usepackage{latexsym}
 \usepackage{ulem}
\usepackage{bm}
\usepackage{epstopdf}
\usepackage{fixltx2e }

\begin{document}

\title[Parametric control of the group velocity]{Subluminal and superluminal light propagation in two kinds of V-type system}

\author{O Budriga}

\address{Laser Department, National Institute for Laser, Plasma and Radiation Physics, P.O. Box MG-36, 077125, Magurele, Romania}
\ead{olimpia.budriga@inflpr.ro}

\begin{abstract}
We study the change of the probe field group velocity from subluminal to superluminal range for two kinds of closed three-level V-type system with spontaneously generated coherence and incoherent pumping field. For the first kind of the V-type system we deduce the analytical formula of the group velocity, while for the second kind of the V-type system we solve numerically the density matrix equations system and apply it to a real system from the LiH molecule. We find that the group index of the probe field can be changed from positive to negative within the appropriate choice of the relative phase between the probe and coupling field phases, the incoherent pumping rate and probe detuning. 
\end{abstract}

\maketitle

\section{Introduction}
In the last decade the investigation of the subluminal light propagation have made a noise in the world thanks to its possible applications on quantnum memory, high-speed optical switches, optical delay lines and optical communication \cite{hau,fleischhauer,phillips,liu,turukhin,ku,mikhailov} and practical applications in optical delay lines \cite{tseng, dahan}. For the first time, Hau {\it et al} observed in a ultracold gas of sodium atoms a subluminal group velocity as low as 17 m/s using the electromagnetically induced transparency (EIT) \cite{hau}. Liu {\it et al} showed, with the aid of EIT, that the light can be stopped in a magnetically trapped, cold cloud of sodium atoms \cite{liu}. Theoretical studies showed that EIT is responsible for the light storage and, therefore, the possibility to use light as quantum memory \cite{fleischhauer}. The same EIT was the tool used to acquire the light storage for times greater than a second in a solid praseodymium doped in Y\textsubscript{2}SiO\textsubscript{5} \cite{longdell}.Wu and Deng reported ultraslow optical solitons in a cold four-state medium \cite{wu}.

Wang {\it et al} used gain-assisted linear anomalous dispersion to demonstrate the superluminal group velocity in atomic caesium gas \cite{wang}, and they showed that the group velocity can exceed the speed of light in vacuum and even become negative. The superluminal or negative group velocity respects the causality being thought of as a pulse-reshaping phenomenon \cite{chiao} and its inverse has a physical significance in connection with the flow energy in light pulses \cite{peatross}, unbraking the relativity theory principle which stipulates that the maximum speed is the light speed {\it c} ($\approx3\cdot10^{8}\;m/s$).  

The control of the group velocity can be accomplished through the electromagnetically induced transparency and electromagnetically induced amplification \cite{kim,bae,bigelow,agarwal,han,joshi}. Based on the $\Lambda$ scheme, with the lower levels coupled by a field, proposed by Agarwal {\it et al} \cite{agarwal}, the experiments carried out on the single atomic transition of the Cs atomic vapour \cite{kim} and Rb atomic vapour \cite{bae} with a standing-wave coupling field have shown the change from subluminal to superluminal group velocity.

The spontaneously generated coherence appears as a result of quantum interference produced by spontaneous decay. For the first time Javanien \cite{Jav} shows the possibility for a $\Lambda$ system with near-degenerated levels to achieve a spontaneously generated coherence (SGC), as a superposition of two receiving states of the spontaneous emission from a single excited state. Effects of the SGC on the group velocity was studied for closed three level $\Lambda$ systems with incoherent pump fields \cite{mahmoudi} and with a strong coupling field, weak probe field and incoherent pump field \cite{dastidar}. Dutta {\it et al} considered a three-level $\Lambda$-type system from the LiH molecule \cite{dastidar}. We will study a three-level V-type system built from other states of LiH molecule than those used in above mentioned paper.   

The subluminal or superluminal group velocity appears in a V-type system with spontaneously generated coherence in the presence of an incoherent pump field \cite{guo} and without it \cite{han,arbiv}. In the case of a weak probe field it was determined the relative phase related to the parameters of the specific system for which the group velocity of the weak probe field is reduced and the probe pulse had undistorted shape \cite{arbiv}. The dependence of the group velocity on the incoherent pumping rate was studied in the presence of the SGC and in its absence, but without phase dependence \cite{guo}. Bai {\it et al} observed a positive group velocity in the absence of the incoherent pump field and a negative group velocity in its presence. The relative phase of the probe and coupling fields can switch the group velocity of the probe field between the subluminal and superluminal range in a V-type system without incoherent pumping field \cite{hanD}. In this paper we apply our general analytical formulas, without constraints on system parameters, from our previous work \cite{Oli2012} to obtain the analytical formula of the probe field group velocity for a closed three level V-type system with spontaneously generated coherence in the presence of an incoherent pumping field. We study also another kind of a three-level V-type system in which the all three fields, the probe, the coupling and the incoherent pumping fields drive both optical allowed transitions. This system was not studied before to the best our knowledge. In this case, the density matrix system equations are solved numerically for a closed three-level V-type system from LiH molecule built using the external field method \cite{ficek}, a real V-type system which was studied by us from the point of view of amplification without population inversion and high refractive index without absorption \cite{OliOptCommun2014}. Consequently our results can be applied to a real molecular system which can be used in an experiment.    
         
A short description of both theoretical systems and their evolution in the density matrix formalism is presented in Section \ref{sec:2}, which has two subsections \ref{sec:2.1} and \ref{sec:2.2}. The Subsection \ref{sec:2.1} is devoted to the three-level V-type system where the probe and coupling fields drive only one transition, while in Subsection \ref{sec:2.2} we present the other kind of three-level V-type system in which the three fields act on both transitions. We achieve the expression of the probe field group velocity in Section \ref{sec:3} for the two kinds of the V-type system in two subsections. The analytical formula of the group velocity for the first kind of a threee-level V-type system is obtained in Subsection \ref{sec:3.1} while Subsection \ref{sec:3.2} is devoted to the formula of the probe field velocity for the V-type system of the second kind. In the Section \ref{sec:4} are presented the numerical results in two subsections. First of them, Subsection \ref{sec:4.1} contains theoretical results from analytical formula of the group velocity and in the second one, Subsection \ref{sec:4.2}, is described the real system from the LiH molecule with numerical results related to the group index. We conclude about our achievements in Section \ref{sec:5}. 

\section{The systems and density-matrix equations}
\label{sec:2}
As we mentioned before we study two kinds of the three-level V-type systems with two very close excited states $\vert1\rangle$ and $\vert2\rangle$ and a ground state $|3\rangle$. In first of them the electric dipole transition moments are chosen
so that one field acts on only one transition. The second one is built as the all three fields, probe, coupling and incoherent pumping fields act on both transitions. All-over this paper we use the SI units.  

\subsection{The V-type system of the first kind}
\label{sec:2.1}
\begin{figure*}
  \includegraphics{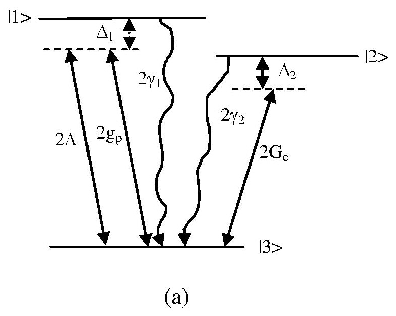}
   \includegraphics{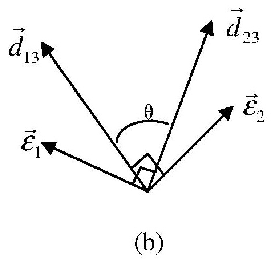}
\caption{(a) The three level V-type system with two near-degenerated excited states $|1\rangle $, $|2\rangle $ and a ground level $|3\rangle $. (b) The electric dipole transition moments are chosen so that one field acts on only one transition. }
\label{fig:1}      
\end{figure*}

The most studied three-level V-type system is drawn schematically in figure \ref{fig:1}(a). The transition $|2\rangle \leftrightarrow |3\rangle $ with frequency $\omega_{23}$ are driven by a coupling field ($\overrightarrow{E}_{2}=\overrightarrow{\epsilon}_{2}e^{-i\omega_{2}t}+c.c.$) with the Rabi frequency $2G_{c}=2\overrightarrow{\epsilon}_{2}\cdot\overrightarrow{d}_{23}/\hbar$. A probe field ($\overrightarrow{E}_{1}=\overrightarrow{\epsilon}_{1}e^{-i\omega_{1}t}+c.c.$) with the Rabi frequency $2g_{p}=2\overrightarrow{\epsilon}_{1}\cdot\overrightarrow{d}_{13}/\hbar$ is applied between the states $\vert1\rangle $ and $|3\rangle $. The transition $|1\rangle\leftrightarrow |3\rangle$ with the frequency $\omega_{13}$ is pumped with a rate $2\Lambda$ by an incoherent field. The detunings of the probe field and the coupling field are $\Delta_{1} = \omega_{13} - \omega_{1}$ and $\Delta_{2} = \omega_{23} - \omega_{2}$, respectively. The rates of spontaneous emission from levels $\vert1\rangle $ and $\vert2\rangle $ to ground level $|3\rangle $ are denoted by $2\gamma_{1}$ and $2\gamma_{2}$, respectively. 

The semiclassical Hamiltonian of this system in the interaction picture, in a rotating-wave frame is written as
\begin{equation}
\label{ec0}
H=-\hbar(g_{p}e^{i\Delta_{1}t}\vert1\rangle\langle3\vert + G_{c}e^{i\Delta_{2}t}\vert2\rangle\langle3\vert + H.c.). 
\end{equation}  

The two excited states, $\vert1\rangle$ and $\vert2\rangle$ must be very close. Consequently the probe field absorption and dispersion depend on the probe and coupling field phases and the existence of the spontaneously generated coherence is related to the complex Rabi frequencies $2g_{p}$ and $2G_{c}$. We write the Rabi frequencies in the form $2g_{p}=2ge^{i\phi_{p}}$ and $2G_{c}=2Ge^{i\phi_{c}}$, where we denoted $\phi_{p}$ and $\phi_{c}$ the phases of the probe field and the coupling field, respectively. The quantities $2g$ and $2G$ are considered real. The system of the density-matrix equations under the rotating-wave approximation, {\it Markov} approximation and dipole approximation, with the phenomenological inclusion of the unidirectional incoherent pump terms and the spontaneous dumping terms, in the steady state case, are
\begin{eqnarray}
\label{ec1}
 \fl-2\gamma_{1}\tilde{\rho}_{11}+2\Lambda\tilde{\rho}_{33}-\eta \rme^{\rmi\phi}\tilde{\rho}_{12}-\eta \rme^{-\rmi\phi}\tilde{\rho}_{21}+\rmi g\tilde{\rho}_{31}-\rmi g\tilde{\rho}_{13}=0\nonumber \\
\fl-2\gamma_{2}\tilde{\rho}_{22}-\eta \rme^{\rmi\phi}\tilde{\rho}_{12}-\eta \rme^{-\rmi\phi}\tilde{\rho}_{21}+\rmi G\tilde{\rho}_{32}-\rmi G\tilde{\rho}_{23}=0\nonumber\\
\fl-(\gamma_{1}+\Lambda+\rmi\Delta_{1})\tilde{\rho}_{13}-\eta \rme^{-\rmi\phi}\tilde{\rho}_{23}+\rmi g(\tilde{\rho}_{33}-\tilde{\rho}_{11})-\rmi G\tilde{\rho}_{12}=0\nonumber \\
\fl-[ \gamma_{1}+\gamma_{2}+\rmi(\Delta_{1}-\Delta_{2})]\tilde{\rho}_{12}+ \rmi g\tilde{\rho}_{32}-\rmi G\tilde{\rho}_{13}\nonumber\\
-\eta \rme^{-\rmi\phi}(\tilde{\rho}_{11}+\tilde{\rho}_{22})=0\nonumber \\
\fl-(\gamma_{2}+\Lambda-\rmi\Delta_{2})\tilde{\rho}_{23}-\eta \rme^{\rmi\phi}\tilde{\rho}_{13} \nonumber\\ -\rmi g\tilde{\rho}_{21}+\rmi G(\tilde{\rho}_{33}-\tilde{\rho}_{22})=0
\end{eqnarray}

The parameter $\eta$ describes the quantum interference between spontaneous emission from excited levels $|1\rangle$ and $|2\rangle$ to ground level $|3\rangle$ and depends on $\theta$, the angle between the two dipole momentum $\overrightarrow{d}_{13}$ and $\overrightarrow{d}_{23}$, $\eta=\eta_{0}\sqrt{\gamma_{1}\gamma_{2}}\cos\theta$, where $\eta_{0}\equiv(\omega_{23}/\omega_{13})^{3/2}$ \cite{cardimona}. The SGC effects are important only for small energy spacing between the two excited levels \cite{Agarwal}, and for high energy spacing such effects will disappear \cite{Paspalakis}. As the excited levels $|1\rangle$ and $|2\rangle$ are near-degenerated, then $\omega_{23}\approx\omega_{13}$ and $\eta_{0}\approx1$. The condition to have spontaneously generated coherence is $\eta\neq0$, which means that we must have a nonorthogonal dipol momentum of the two transitions. Therefore we choose the dipole momentum so that one field acts on one transition ($\overrightarrow{d}_{13}\perp\overrightarrow{\epsilon}_{2}$, $\overrightarrow{d}_{23}\perp\overrightarrow{\epsilon}_{1}$) as it can be shown in figure 1(b). Rabi frequencies are connected to the angle $\theta$ by the relations $2g_{p}=2|\overrightarrow{E}_{1}||\overrightarrow{d}_{13}|\sin\theta/\hbar$ and $2G_{c}=2|\overrightarrow{E}_{2}||\overrightarrow{d}_{23}|\sin\theta/\hbar$. The influence of the angle $\theta$ on the probe coefficient absorption and refractive index was evidenced in our paper \cite{Oli2013}. The nonorthogonality of the dipolar momentum can be achieved from the mixing of the levels arising from internal \cite{physrevlet77} or external fields \cite{hakuta, faist, berman, patnaik, optcomun179}. 

The states $|2\rangle$ and $|3\rangle$ are dressed by the strong coupling field. The dressed states are written as 
\begin{eqnarray}
\label{ec13bis}
|+\rangle = \cos\alpha|2\rangle + \sin\alpha|3\rangle\nonumber\\  
|-\rangle  = -\sin\alpha|2\rangle + \cos\alpha|3\rangle,
\end{eqnarray}
where $0 \le 2\alpha < \pi$, $tg(2\alpha) = -2G/\Delta_{2}$. The eigenvalues corresponding to the two dressed states have the expresions $\hbar\lambda_{+}=\hbar(\Delta_{2}+\sqrt{\Delta_{2}^{2}+4G^{2}})/2$ and $\hbar\lambda_{-}=(\Delta_{2}-\sqrt{\Delta_{2}^{2}+4G^{2}})/2$, respectively. This type of the dressed states will be used for the building of a three-level V-type system of the second type from the real system of LiH molecule.

We have derived the analytical solution of the system (\ref{ec1}) in the general case without restrictions on the probe and coupling field, in all order of $g$ and $G$ in our previous article \cite{Oli2012} and we shall use below.

\subsection{The V-type system of the second kind}
\label{sec:2.2}
There is the possibility that the probe, the coupling and the incoherent pumping fields act on both transtions from the three-level V-type system described in the previous section, the transitions between the excited levels $|1\rangle$ and $|2\rangle$ to ground level $|3\rangle$. In this case appear two additional Rabi frequencies of the probe and coupling fields, $2g_{p}'=2\overrightarrow{E}_{1}\cdot\overrightarrow{d}_{23}/\hbar$ and $2G_{c}'=2\overrightarrow{E}_{2}\cdot\overrightarrow{d}_{13}/\hbar$, respectively. Assuming that the two laser angular frequencies have almost the same values, i.e. $\omega_{1}\approx\omega_{2}$ the semiclassical Hamiltonian of this system in the interaction picture, in a rotating-wave frame will be
\begin{equation}
\label{ec2}
H=-\hbar[(g_{p}+G_{c}')e^{i\Delta_{1}t}\vert1\rangle\langle3\vert +(G_{c}+g_{p}')e^{i\Delta_{2}t}\vert2\rangle\langle3\vert + H.c.]. 
\end{equation}  

Using the same formalism as in the previous subsection we obtain the system of equations for the density matrix elements as  
$$-2\gamma_{1}\tilde{\rho}_{11}+2\Lambda\tilde{\rho}_{33}-\eta e^{i\phi}\tilde\rho_{12}-\eta e^{-i\phi}\tilde\rho_{21}+i(g+G')\tilde\rho_{31}-i(g+G')\tilde\rho_{13}=0$$
$$-2\gamma_{2}\tilde{\rho}_{22}+2\Lambda\tilde{\rho}_{33}-\eta e^{i\phi}\tilde{\rho}_{12}-\eta e^{-i\phi}\tilde{\rho}_{21}+i(G+g')\tilde{\rho}_{32}-i(G+g')\tilde{\rho}_{23}=0$$
$$-(\gamma_{1}+2\Lambda+i\Delta_{1})\tilde{\rho}_{13}-\eta e^{-i\phi}\tilde{\rho}_{23}+i(g+G')(\tilde{\rho}_{33}-\tilde{\rho}_{11})-i(G+g')\tilde{\rho}_{12}=0$$
$$-(\gamma_{2}+2\Lambda-i\Delta_{2})\tilde{\rho}_{23}-\eta e^{i\phi}\tilde{\rho}_{13}- \nonumber\\
i(g+G')\tilde{\rho}_{21}+i(G+g')(\tilde{\rho}_{33}-\tilde{\rho}_{22})=0$$
$$-[\gamma_{1}+\gamma_{2}+i(\Delta_{1}-\Delta_{2})]\tilde{\rho}_{12}-\eta e^{-i\phi}(\tilde{\rho}_{11}+\tilde{\rho}_{22})+$$
\begin{equation}
\label{ec3}
i(g+G')\tilde{\rho}_{32}-i(G+g')\tilde{\rho}_{13}=0
\end{equation}
with {\it g, g', G, G'} the real Rabi frequencies corresponding to the complex Rabi frequencies $g_{p}$, $g_{p}'$, $G_{c}$ and $G_{c}'$, respectively. This system of linear equations differs from the system of linear equations (\ref{ec1}) by the appearance of the term $2\Lambda\tilde{\rho}_{33}$ in the second equation. The other equations have the same terms as (\ref{ec1}), where the $\Lambda$, $g$ and $G$ are replaced by $2\Lambda$, $g+G'$ and $G+g'$, respectively.

We shall use the model proposed by Ficek and Swain to develop a Vee-type system of the second kind with antiparallel dipole moments, which consist in appling a strong laser field to one of the two transitions in a Lambda-type system \cite{ficek}. We chose the real three level $\Lambda$ system from LiH molecule and the data from the paper of Bhattacharjee {\it et al} \cite{anindita}. 

\section{The group velocity}
\label{sec:3}
The group velocity $v_{\rm g}$ of an electromagnetic field is defined as
\begin{eqnarray}
\label{ec4}
\fl v_{\rm g}=\frac{\rmd\omega}{\rmd k}=\frac{c}{n+\omega\frac{\rmd n}{\rmd\omega}},
\end{eqnarray}
where $\omega$ is the field frequency, $k$ is the wave number, {\it n} is the refractive index of the medium and {\it c} is the vacuum light speed. In a dilute medium the index of refraction is related to the real part of dielectric susceptibility of the medium $Re\chi_{e}$ by the relation $n=1+Re\chi_{e}/2$ in SI. Consequently the group velocity depends on the dielectric susceptibility 
\begin{eqnarray}
\label{ec5}
\fl v_{\rm g}=\frac{c}{1+\frac{1}{2}Re\chi_{e}+\frac{\omega}{2}\frac{\rmd Re\chi_{e}}{\rmd\omega}}.
\end{eqnarray}
We can use a more confortable quantity, named group index $n_{\rm g}=c/v_{\rm g}$. If $n_{\rm g} > 1$ then group velocity is lower than the vacuum speed velocity and the field has a subluminal group velocity. Otherwise, $n_{\rm g} < 1$, and the field has a superluminal group velocity. 

\subsection{The analytical formula in the case of the V-type system of the first kind}
\label{sec:3.1}
For the three level V-type system of the first kind the susceptibility of the medium at the probe field is proportional with the density matrix element $\tilde\rho_{31}$
\begin{eqnarray}
\label{ec6}
\fl \chi_{\rm e}=\frac{2Nd_{31}^2}{\hbar\epsilon_{0}g}\tilde\rho_{31},
\end{eqnarray}  
where $N$ is the density of the V-type systems and $\epsilon_{0}$ is the vacuum dielectric permittivity. From relations (\ref{ec3}), (\ref{ec4}) and our prior analytical results of the density matrix element $\tilde\rho_{31}$ \cite{Oli2012}, we derive the quantity $n_{\rm g}-1$ 
 \begin{eqnarray}
\label{ec7}
\fl n_{\rm g}-1=\frac{N d_{31}^2}{\hbar\epsilon_{0}g}\lbrack Re\tilde\rho_{31}-\omega_{1}\frac{\rmd Re\tilde\rho_{31}}{\rmd \Delta_{1}}\rbrack.
\end{eqnarray}
Starting from the expression (4) of $Re\tilde{\rho}_{31}$ and the relations which are detailed in the Appendix of the previous paper \cite{Oli2012} is straightforward to obtain the analytical expression of $\rmd Re\tilde{\rho}_{31}/\rmd \Delta_{1}$. The exhaustive formula of the derivative of the $Re\tilde{\rho}_{31}$ with respect to the probe detuning $\Delta_{1}$ is written in the Appendix of this work.

\subsection{The expression in the case of the V-type system of the second kind}
\label{sec:3.2}
The polarization of the medium which consists in three-level V-type systems of the second kind, by the probe field is
\begin{eqnarray}
\label{ec8}
\overrightarrow{P}(\omega_{1})&=NTr(\hat{\tilde\rho}\hat{\overrightarrow{d}})\nonumber\\
&=2N(\tilde\rho_{13}\overrightarrow{d}_{31}+\tilde\rho_{23}\overrightarrow{d}_{32})
\end{eqnarray}
where {\it Tr} represents the trace and $N$ is the density of the V-type systems of the second type. Also, the polarization of the medium by the probe field, $\overrightarrow{P}(\omega_{1})$ is directly proportional with the electric susceptibility of the medium $\chi_{e}$ and the electric probe field $\overrightarrow{E}_{1}$
\begin{eqnarray}
\label{ec9}
\overrightarrow{P}(\omega_{1})=\epsilon_{0}\chi_{e}(\omega_{1})\overrightarrow{E}_{1}.
\end{eqnarray}
For the three-level V-type system that we will obtain in the LiH molecule, the electric dipoles $\overrightarrow{d}_{13}$ and $\overrightarrow{d}_{23}$ are antiparallel. The fact that the probe and coupling fields drive and couple both transitions $\vert 1\rangle\rightarrow|3\rangle$ and $\vert 2\rangle\rightarrow|3\rangle$ implies that there are no restrictions over the field polarizations and allow us to choose the electric polarizations $\overrightarrow{\epsilon}_{1}$ parallel with the dipole transition moment $\overrightarrow{d}_{13}$ and $\overrightarrow{\epsilon}_{2}$ parallel with the dipole transition moment $\overrightarrow{d}_{23}$. Therefore, the electric susceptibility becomes
\begin{eqnarray}
\label{ec10}
\chi_{e}=\frac{2N}{\epsilon_{0}E_{1}}(\tilde\rho_{13}{d}_{31}-\tilde\rho_{23}{d}_{32}).
\end{eqnarray}      
With this choice of the electric field polarizations the real Rabi frequencies have the expressions $2g=2E_{1}d_{13}/\hbar$, $2g'=-2E_{1}d_{23}/\hbar$, $2G=2E_{2}d_{23}/\hbar$ and $2G'=-2E_{2}d_{13}/\hbar$. From the relations (\ref{ec5}) and (\ref{ec10}) the group index of the probe field can be written
\begin{eqnarray}
\label{ec11}
\fl n_{\rm g}-1=\frac{Nd_{13}}{\epsilon_{0}\hbar g}\lbrace d_{13}[Re\tilde\rho_{13}-(\omega_{13}-\Delta_{1})\frac{\rmd Re\tilde\rho_{13}}{\rmd \Delta_{1}}]\nonumber\\
-d_{23}[Re\tilde\rho_{23}-(\omega_{13}-\Delta_{1})\frac{\rmd Re\tilde\rho_{23}}{\rmd \Delta_{1}}]\rbrace.
\end{eqnarray} 
We will use the above formula in our numerical calculations and will present and discuss the results in the Subsection \ref{sec:4.2}.

\section{Numerical results}
\label{sec:4}
We do numerical calculations of the group index $n_{\rm g}-1$ related to the parameters of a theoretical three-level V-type system of the first kind by using the analytical formulas (\ref{ec7}) and (\ref{a1}). We achieve the values of the relative phase, incoherent pumping rate and probe detuning for which the group velocity of the probe field is lower or greater than the light speed in vacuum. A real three-level V-type system of the second kind obtained from the three vibrational levels of the LiH moecule is investigated to find how the group velocity of the probe field can be changed from subluminal to superluminal.

\subsection{The V-type system of the first kind}
\label{sec:4.1}
\begin{figure*}
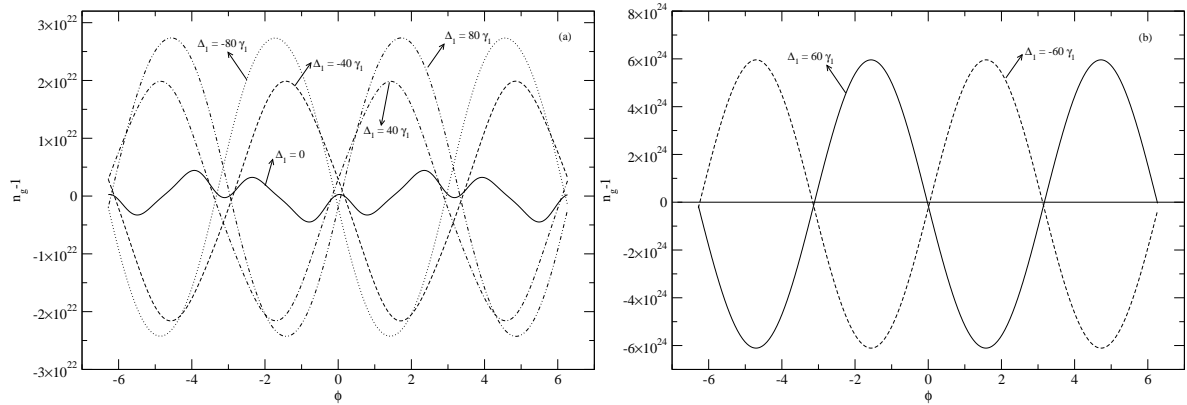

 \resizebox{\columnwidth}{!}{
  \includegraphics{Fig2a.eps}
   \includegraphics{Fig2b.eps}}
\caption{The group index of the probe field $n_{\rm g}-1$ via relative phase $\phi$ of the V-type system of the first kind with the parameters $\gamma_{1} = 1 Hz$, $\gamma_{2} = 0.4\gamma_{1}$, $\Lambda = 0.5\gamma_{1}$, $\Delta_{2} = 0.01\gamma_{1}$, $g = 0.04\gamma_{1}$, $G = 60\gamma_{1}$, $\omega_{1} = 10^{11}\gamma_{1}$ and (a) $\Delta_{1} = 0, \pm40 \gamma_{1}, \pm80 \gamma_{1}$, (b) $\Delta_{1} =\pm60 \gamma_{1}$. }
\label{fig2}      
\end{figure*}

In our numerical calculations we considered a three-level V-type system of the first kind with the parameters $\gamma_{1} = 1 $Hz, $\gamma_{2} = 0.4\gamma_{1}$, $\Delta_{2} = 0.01\gamma_{1}$, $g = 0.04\gamma_{1}$, $G = 60\gamma_{1}$, $\omega_{1} = 10^{11}\gamma_{1}$ and $\theta = \pi$. In figure \ref{fig2} are plotted the graphics of the group index $n_{g}-1$ versus the relative phase $\phi$ for some probe field detunings $\Delta_{1}$ and $\Lambda = 0.5\gamma_{1}$. The change of the relative phase $\phi$ leads to the periodically transition of the group velocity from the subluminal ($n_{g}-1 > 0$) to superluminal ($n_{g}-1 < 0$) regimen. The graphics for the probe field detuning with opposite signs are dephased with $\pi$. 

From figure \ref{fig2}(a) we observe that the highest values of the index group are obtained for the detuning of the probe field given by the frequency formula of the dressed state $|+\rangle$ from the relation (\ref{ec13bis}), $\Delta_{1} = (\Delta_{2}+\sqrt{\Delta_{2}^{2}+4G^{2}})/2 = 60.005\gamma_{1}$. The group index $n_{g}-1$ for the probe field detuning $\Delta_{1} = 60\gamma_{1}$ is of the order $10^{24}$ while for the other values of the probe field detuning is lower with two order of magnitude. For this reason we shall choose in the next calculations the probe field detuning as $60\gamma_{1}$.  

\begin{figure}
  \resizebox{0.75\columnwidth}{!}{
  \includegraphics{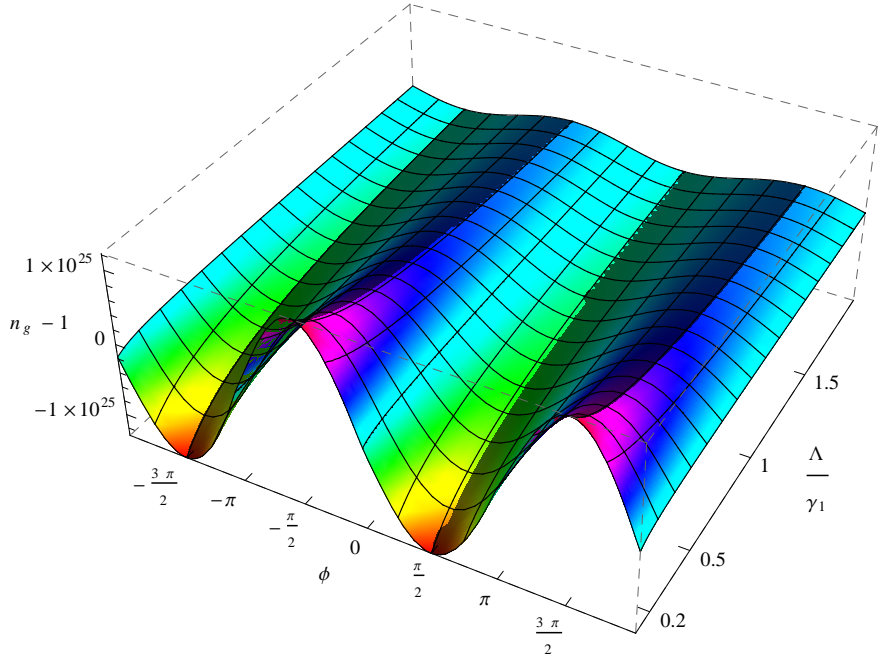}
  }
\caption{Three-dimensional plot of the group index $n_{g}-1$ (vertical axis) versus relative phase $\phi$ of the probe and coupling fields (horizontal axis across the page) and the incoherent pumping field $\Lambda/\gamma_{1}$ (horizontal axis into the page) with the other parameters same as in figure \ref{fig2}.}
\label{fig3}      
\end{figure}

As we mentioned in the introduction, our goal is to find the combined effect of the relative phase and the incoherent pumping rate on the group velocity of the probe field. The figure \ref{fig3} emphasizes the behaviour of the group index $n_{\rm g}-1$ when the relative phase $\phi$ and the incoherent pumping rate $\Lambda$ vary from zero to $2\gamma_{1}$. 

\begin{figure}
  \resizebox{0.75\columnwidth}{!}{
  \includegraphics{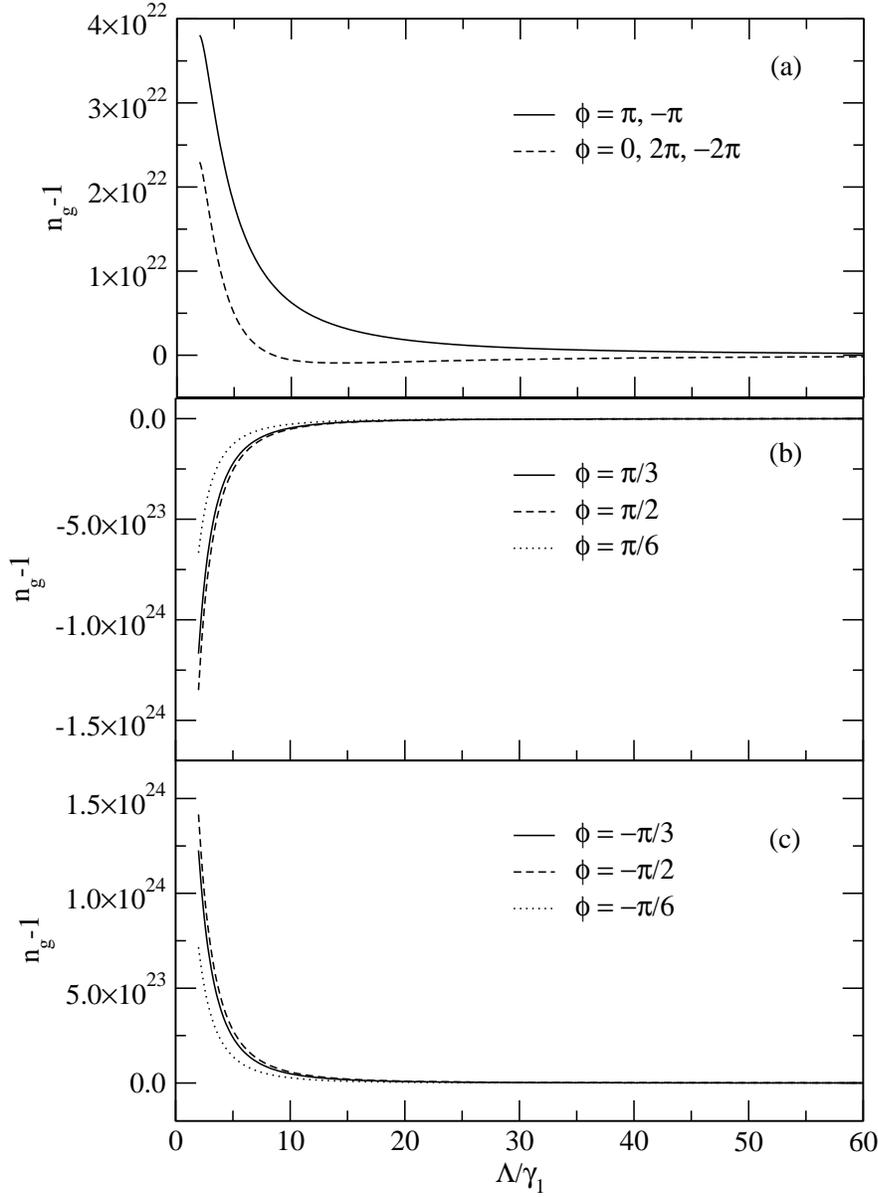}
  }
\caption{The group index $n_{g}-1$ versus the incoherent pumping field $\Lambda/\gamma_{1}$ for (a) $\phi = 0, \pm\pi, \pm2\pi$, (b) $\phi = \pi/6, \pi/3, \pi/2$ and (c) $\phi = -\pi/6, -\pi/3, -\pi/2$. The other parameters are same as in figure \ref{fig2}.}
\label{fig4}      
\end{figure}

For any value of the incoherent pumping rate of the probe field $\Lambda$ between 0 and $2\gamma_{1}$ the probe field exhibits a group velocity lower than the speed of light $c$ or higher than $c$ with the period $2\pi$ of the relative phase $\phi$. At the same time with the increase of the incoherent pumping rate of the probe field $\Lambda$ the highest positive values of the group index $n_{g}-1$, obtained for relative phase $\phi = 3\pi/2$ decrease and the lowest negative values, obtained for $\phi = \pi/2$ increase. The same behaviour can be seen for an extended domain for the incoherent pumping rate $\Lambda$ from $2\gamma_{1}$ to $60\gamma_{1}$ in the figure \ref{fig4}. 

The group velocity of the probe field $v_{g}$ is superluminal for any incoherent pumping rate $\Lambda$ and relative phase $0 < \phi < \pi$ and subluminal for $\pi < \phi < 2\pi$. With the increase of the incoherent pumping rate $\Lambda$ and $\phi$ in the range $0 < \phi < \pi$ the group index $n_{g}-1 = c/v_{g} - 1$ increases, that means the group velocity of the probe field $v_{g}$ decreases. In the other domain of the relative phase $\pi < \phi < 2\pi$ the group index $n_{g}-1 = c/v_{g} - 1$ decreases and correspondingly the group velocity $v_{g}$ increases. In the case of the relative phase $\phi = \pi$ for incoherent pumping rate $\Lambda \le 0.9\gamma_{1}$ the group velocity of the probe field is superluminal ($n_{g}-1 < 0$) and for the incoherent pumping field $\Lambda \ge \gamma_{1}$ the group velocity $v_{g}$ is subluminal ($n_{g}-1 > 0$). For the relative phase $\phi = 2k\pi$, $k$ integer number, the group velocity of the probe field is higher than the velocity of the light $c$ when the incoherent pumping rate $\Lambda \le \gamma_{1}$ and $\Lambda > 8.4\gamma_{1}$. The figures \ref{fig3} and \ref{fig4} shows us, also, for the incoherent pumping rate $\gamma_{1} <\Lambda \le 8.4\gamma_{1}$ and relative phase $\phi = 2k\pi$, $k$ integer number, one has $v_{g} < c$. We can observe from the figures \ref{fig4} (b) and (c) that the group index is oposite in sign for relative phase oposite in sign, i.e. $n_{g}(-\phi) = -n_{g}(\phi)$. Therefore, the group index is an odd function in relative phase $\phi$, for any $\phi \ne 2k\pi$, where {\it k} is an integer number.

\subsection{The V-type system of the second kind in LiH molecule}
\label{sec:4.2}
As we mentioned in the Subsection \ref{sec:3.2} we will solve the system of equations (\ref{ec3}) and will use the relation (\ref{ec11}) for the group index $n_{g}-1$ in the case of the V-type system of the second kind in LiH molecule, which we have described in our previously work \cite{OliOptCommun2014}.   

\begin{figure*}
  \includegraphics{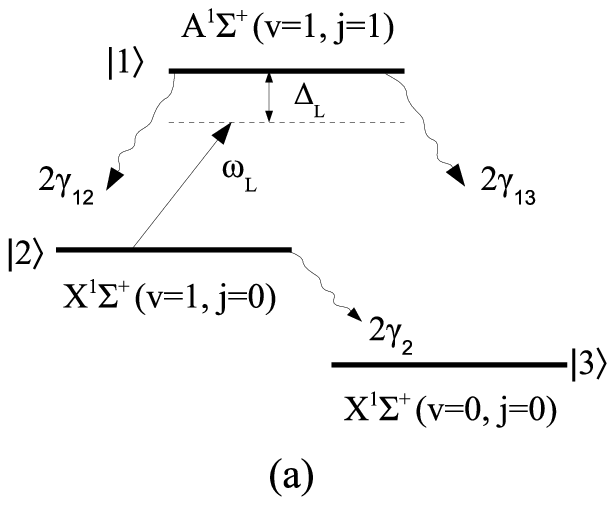}
  \includegraphics{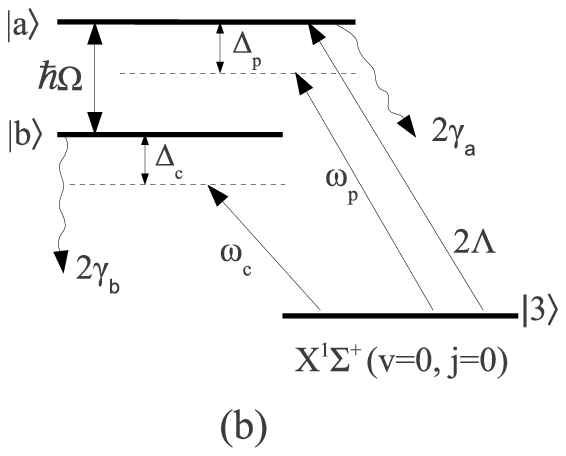}
\caption{(a) LiH molecule $\Lambda$ system consisted of the excited level $A^{1}\Sigma^{+} (v=1,\:j=1)$ and vibrational states $X^{1}\Sigma^{+} (v=1,\:j=0)$ and $X^{1}\Sigma^{+} (v=0,\:j=0)$ of the ground level. An external strong laser field couples the two excited levels named $\vert 1\rangle$ and $\vert 2\rangle$. (b) The subsystem with the upper dressed states $\vert a\rangle$ and $\vert b\rangle$ and the ground level  $X^{1}\Sigma^{+} (v=0,\:j=0)$ form a three-level Vee-type system with the dipole moments of the two transitions,$\overrightarrow{d}_{a3}$ and $\overrightarrow{d}_{b3}$, antiparallel.} 
\label{fig5}      
\end{figure*}

Let us apply a strong field between $A^{1}\Sigma^{+} (v=1,\:j=1)$ and $X^{1}\Sigma^{+} (v=1,\:j=0)$ states of the LiH molecule (see figure \ref{fig5} (a)). It then produces the dressed states
\begin{eqnarray}
\label{ec20}
 \vert a\rangle=\sin{\psi}\vert2\rangle+\cos{\psi}\vert1\rangle\nonumber\\
 \vert b\rangle=\cos{\psi}\vert2\rangle-\sin{\psi}\vert1\rangle    
\end{eqnarray}     
with $0 \le 2\psi < \pi$, $tg(2\psi)=-2G_{L}/\Delta_{L}$, where $G_{L}$ is the real Rabi frequency of the strong coupling field and $\Delta_{L}$ the detuning of the strong laser frequency from the molecular transition $\vert1\rangle\rightarrow|2\rangle$ \cite{Cohen}. The separation energy between the two dressed states $\vert a\rangle$ and $\vert b\rangle$ are equal with $\hbar\Omega$, where $\Omega=\sqrt{\Delta_{L}^{2}+4G_{L}^{2}}$. 

To build a V-type system of the second kind these two dressed states ($\vert a\rangle$ and $\vert b\rangle$) are used as two upper levels and coupled to the ground level $X^{1}\Sigma^{+} (v=0,\:j=0)$ by probe and strong coupling fields, respectively (see figure \ref{fig5}). We find that the dipole matrix elements between the dressed states and the ground level are $\overrightarrow{d}_{a3}=\overrightarrow{d}_{13}\cos{\psi}$ and $\overrightarrow{d}_{b3}=-\overrightarrow{d}_{13}\sin{\psi}$, where $\overrightarrow{d}_{13}$ is the dipole transition moment between $A^{1}\Sigma^{+} (v=1,\:j=1)$ and $X^{1}\Sigma^{+} (v=0,\:j=0)$ states. Thus, the two dressed states $\vert a\rangle$, $\vert b\rangle$ and the ground state $X^{1}\Sigma^{+} (v=0,\:j=0)$ bahave as a V-type system with antiparallel electric dipole transition moments. We consider that the weak probe laser, coupling laser and incoherent pumping laser act on both transitions between the dressed states $\vert a\rangle$, $\vert b\rangle$ and the ground state $\vert 3\rangle$. We keep the notations from our paper \cite{OliOptCommun2014}. 

For this three-level V-type system of the second kind from the relation (\ref{ec11}) the group index of the probe field $n_{\rm g}-1$ in SI units becomes
\begin{eqnarray}
\label{ec23}
\fl n_{\rm g}-1=\frac{Nd_{a3}}{\epsilon_{0}\hbar g}\lbrace d_{a3}[Re\tilde\rho_{a3}-(\omega_{a3}-\Delta_{p})\frac{\rmd Re\tilde\rho_{a3}}{\rmd \Delta_{p}}]\nonumber\\
-d_{b3}[Re\tilde\rho_{b3}-(\omega_{a3}-\Delta_{p})\frac{\rmd Re\tilde\rho_{b3}}{\rmd \Delta_{p}}]\rbrace.
\end{eqnarray}

\begin{figure}
  \resizebox{0.75\columnwidth}{!}{
  \includegraphics{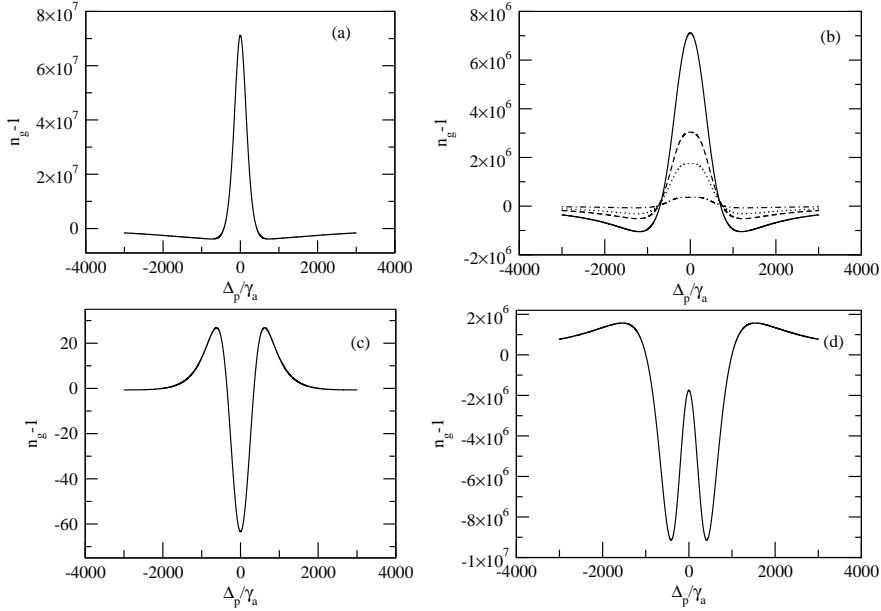}
  }
\caption{Group index $n_{g}-1$ versus the probe detuning $\Delta_{p}/\gamma_{a}$ for the incoherent pumping rate (a) $\Lambda = 0$, (b) $\Lambda = 0.5 \gamma_{a}$ (solid line), $0.7 \gamma_{a}$ (dashed line), $0.8 \gamma_{a}$ (dotted line), $0.95 \gamma_{a}$ (dot-dash line), (c) $\Lambda = \gamma_{a}$ and (d) $\Lambda = 10 \gamma_{a}$. The other parameters of the three-level V-type system of the second kind from the LiH molecule are $\gamma_{a} = 2.475\cdot10^{4}$ Hz, $\gamma_{b} = 0.33\gamma_{a}$, $g = 0.033\gamma_{a}$, $g' = -0.019\gamma_{a}$, $G = 235.35\gamma_{a}$, $G' = -407.64\gamma_{a}$, $\phi = \pi$, $\cos{\psi} = 0.5$, $N = 10^{12}$ molecules/cm\textsuperscript{3}.}
\label{fig6}      
\end{figure}

In figure \ref{fig6} are drawn the graphics representing the dependence of the group index $n_{g}-1$ on the probe detuning $\Delta_{p}/\gamma_{a}$ for different values of the incoherent pumping rate $\Lambda = 0, 0.5 \gamma_{a}$, $0.7 \gamma_{a}$, $0.8 \gamma_{a}$, $0.95 \gamma_{a}$, $\gamma_{a}$ and $10 \gamma_{a}$ and the parameters of the system $\gamma_{a} = 2.475\cdot10^{4}$ Hz, $\gamma_{b} = 0.33\gamma_{a}$, $g = 0.033\gamma_{a}$, $g' = -0.019\gamma_{a}$, $G = 235.35\gamma_{a}$, $G' = -407.64\gamma_{a}$, $\phi = \pi$, $\cos{\psi} = 0.5$, $N = 10^{12}$ molecules/cm\textsuperscript{3}. With $\gamma_{a}$ we denoted the spontaneous decay rate of the dressed state $a$ and with $\gamma_{b}$ the spontaneous decay rate of the dressed state $b$.

It can be seen that for probe field detunings no so higher than the Rabi frequency $G = 235.35\gamma_{a}$ the group velocity of the probe field is subluminal for the incoherent pumping rate $\Lambda < \gamma_{a}$ and superluminal for $\Lambda \ge \gamma_{a}$. At the values of the probe field detuning $\Delta_{p}$ very far with the Rabi frequency $G$ the behaviour of the group velocity is inversely. We choose the probe field detuning $\Delta_{p}$ equal with the Rabi frequency $G$, same as in our previous paper \cite{OliOptCommun2014}. 

\begin{figure}
  \resizebox{0.75\columnwidth}{!}{
  \includegraphics{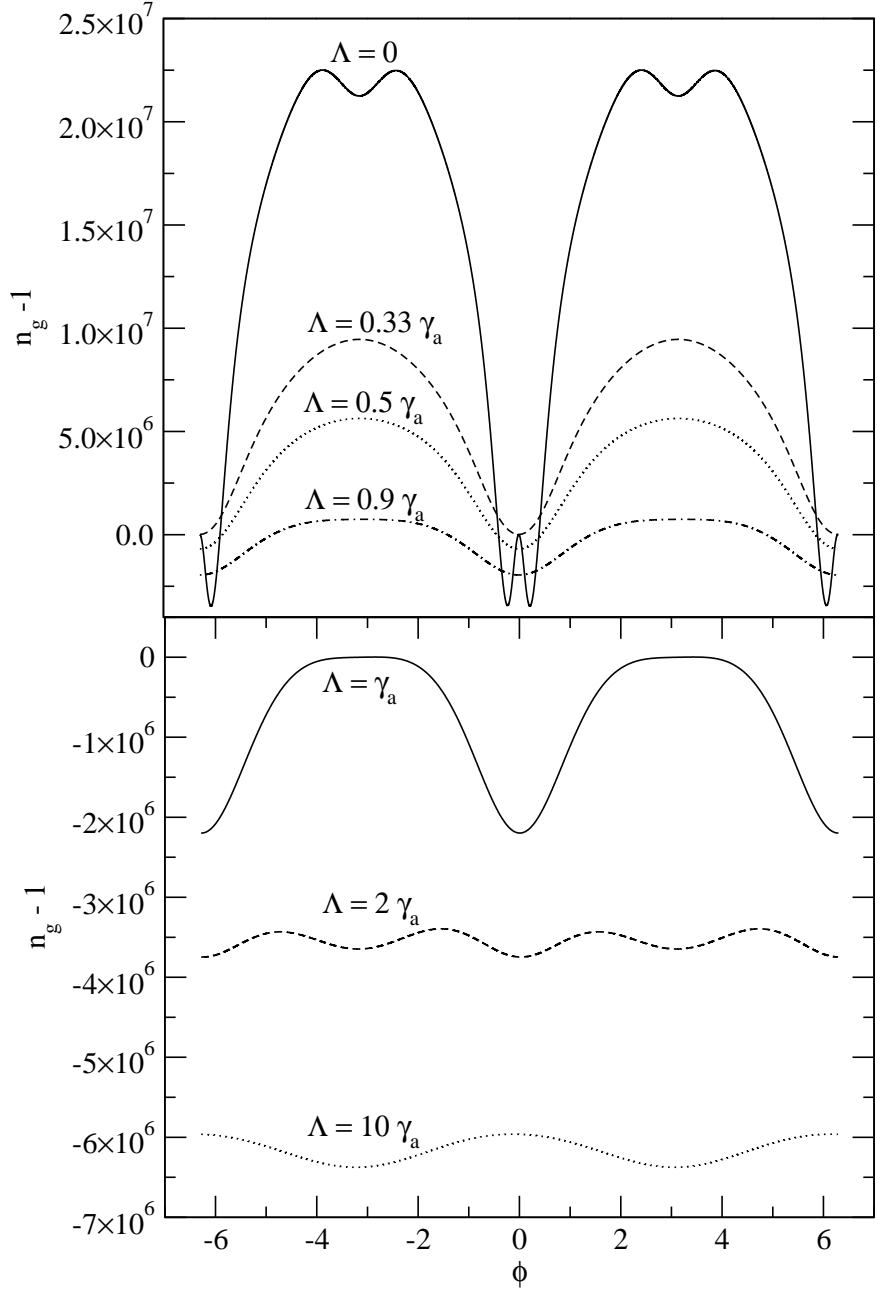}
  }
\caption{Group index $n_{g}-1$ versus relative phase $\phi$ between the probe and coupling fields for different incoherent pumping rate $\Lambda$. The other system parameters are same as in figure \ref{fig6} and $\Delta_{p} = G$.}
\label{fig7}      
\end{figure}

The variation with the relative phase $\phi$ of the group index of the probe field $n_{g}-1$ is shawn in the figure \ref{fig7} for some values of the incoherent pumping rate $\Lambda$: $0, 0.33 \gamma_{a},0.5 \gamma_{a}, 0.9 \gamma_{a}$ (see figure \ref{fig7} (a)), $\gamma_{a}, 2 \gamma_{a}$ and $10 \gamma_{a}$ (see figure \ref{fig7} (b)).

The group index $n_{g}-1$ of the probe field varies periodically with the relative phase $\phi$, with the period $2\pi$, same as for the three-level V-type system of the first kind. In the case of the incoherent pumping rate $\Lambda \le 0.33\gamma_{a}$ the group index is positive, that means the group velocity of the  probe field is lower than the speed of light $c$ for any relative phase $\phi$ as can be seen in figure \ref{fig7} (a). At the same time if the incoherent pumping rate $\Lambda > \gamma_{a}$, then the group index is negative and the group velocity is higher than the speed of light in vacuum $c$, for any relative phase (see figure \ref{fig7} (b)). For the intermediate values of the incoherent pumping rate $\Lambda$ the group index of the probe field $n_{g} - 1$ is positive or negative, dependent on $\phi$. That means the group velocity changes from subluminal to superluminal along with the variation of the relative phase.   

\begin{figure}
  \resizebox{0.75\columnwidth}{!}{
  \includegraphics{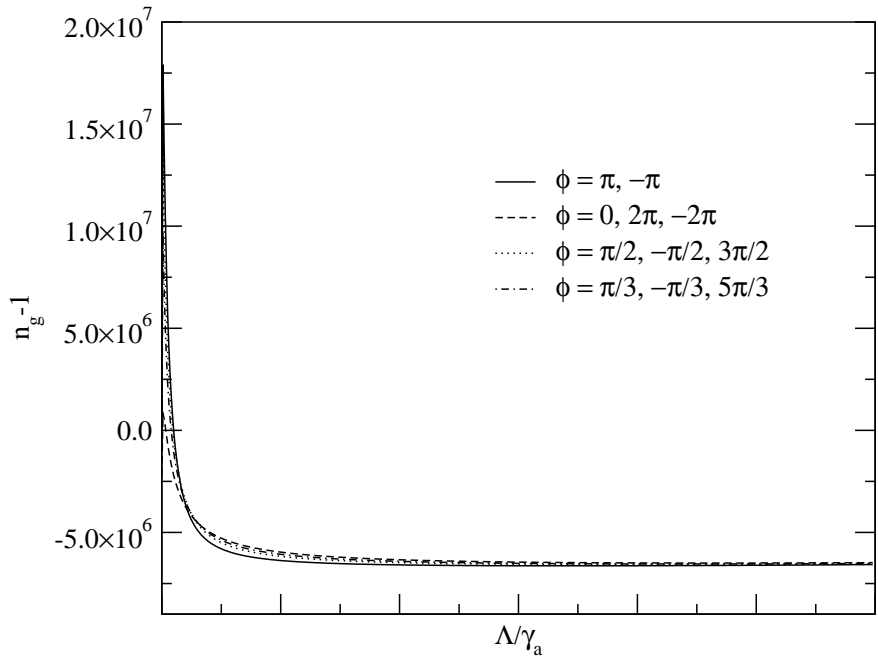}
  }
\caption{Group index $n_{g}-1$ versus incoherent pumping rate $\Lambda/\gamma_{a}$ for some relative phase $\phi$. The other system parameters are same as in figure \ref{fig7}.}
\label{fig8}      
\end{figure}

In figure \ref{fig8} is illustrated the dependence of the group index on the incoherent pumping rate $\Lambda$, for some values of the relative phase $\phi$.    

When the incoherent pumping rate increases the group index decreases, and consequently the group velocity increases. This behaviour of the group index is the same for any value of the relative phase $\phi$. Also, the graphics are the same for the relative phase with opposite sign, i. e. $(n_{g}-1)(\phi)=(n_{g}-1)(-\phi)$. Hence, the group index is an even function in $\phi$. This feature for $\phi \ne 2k\pi$, where {\it k} is an integer number, is distinct on the case of the three-level V-type system of the first kind, when the group index is an odd function in relative phase $\phi$. One can remark the period of $2\pi$ in relative phase $\phi$ of the group index, i. e. $(n_{g}-1)(\phi+2\pi)=(n_{g}-1)(\phi)$. As examples, the plots of the group index from the figure \ref{fig8} coincide for relative phase $\phi$ equals with 0 and $2\pi$, $\pi$ and $-\pi$, $-\pi/2$ and $3\pi/2$ and for $-\pi/3$ and $5\pi/3$.  

The V-type system of the second kind in the heterogeneous molecule LiH is a real example of a system in which the probe field can achive both subluminal and superluminal group velocity. 

\section{Conclusions}
\label{sec:5}
We studied two kinds of the three-level closed V-type systems. In first of them the probe, coupling and incoherent pumping fields act only on one transition, while in the V-type system of the second kind the all fields act on both optical allowed transitions. For the V-type system of the first kind we considered a theoretical system and for the V--type system of the second kind we analyzed a real system from the heterogeneous LiH molecule. We found that in both kinds of the three-level V-type systems the proper choice of the relative phase between the probe and coupling fields and the incoherent pumping rate leads to the subluminal or superluminal propagation of the probe field. The group index of the probe field varies priodically with relative phase $\phi$, being higher or lower than the speed of light in vacuum $c$. The probe field changes its group velocity from subluminal to superluminal periodically with the  variation of the relative phase $\phi$, for any value of the incoherent pumping rate $\Lambda$ in the case of the V-type system of the first kind. The periodicity with the relative phase of the group index of the probe field remains the same for the V-type system of the second kind from LiH molecule, but the probe field exhibits a subluminal or superluminal group velocity depending on the relative phase only for the incoherent pumping rate $\Lambda$ lower than the spontaneous decay rate of the highest dressed state $\gamma_{a}$ and higher than the value $0.33 \gamma_{a}$. When the incoherent pumping rate exceeds the value $\gamma_{a}$ the probe field has a superluminal group velocity and when the incoherent pumping rate is lower than the value $0.33 \gamma_{a}$ the probe field has a subluminal group velocity, for any value of the relative phase. The V-type system from the LiH molecule is a real system which can be used in experiments to obtain subluminal or superluminal group velocity of the probe field, using the relative phase and the incoherent pumping control.

\ack

The author thanks the support of Ministry of Education and Research, Romania (program Laplas 3, PN 09 39).

\appendix
\section*{Appendix}
\setcounter{section}{1}
Using the expression of the real part of the density matrix element $\tilde{\rho}_{31}$, relation (4) from \cite{Oli2012}, we obtain the formula of the derivative with respect to the probe field detuning $\Delta_{1}$ as
\begin{eqnarray}
\label{a1}
\fl \frac{\rmd Re\tilde{\rho}_{31}}{\rmd \Delta_{1}}=\frac{M}{N},
\end{eqnarray}
where the numerator M is
\begin{eqnarray}
\label{a2}
\fl M=A^{2}(b_{1}'c_{1}-b_{1}c_{1}')+AC(b_{1}'c_{7}-b_{1}c_{7}'+b_{2}'c_{1}-b_{2}c_{1}')
\fl \nonumber\\+AD(b_{1}'c_{4}-b_{1}c_{4}'+b_{3}'c_{1}-b_{3}c_{1}')+CD(b_{2}'c_{4}-b_{2}c_{4}'+b_{3}'c_{7}-b_{3}c_{7}')
\fl\nonumber\\+(A'C-AC')(b_{1}c_{7}-b_{2}c_{1})+(A'D-AD')(b_{1}c_{4}-b_{3}c_{1})
\fl\nonumber\\+C^{2}(b_{2}'c_{7}-b_{2}c_{7}')+(AC'-A'C)(b_{2}c_{1}-b_{1}c_{7})
\fl\nonumber\\+(C'D-CD')(b_{2}c_{4}-b_{3}c_{7})+D^{2}(b_{3}'c_{4}-b_{3}c_{4}')
\end{eqnarray}
and prime means the derivative with respect to the probe field detuning $\Delta_{1}$. The denominator $N$ is
\begin{eqnarray}
\label{a3}
\fl N=(c_{1}A+c_{7}C+c_{4}D)^{2},
\end{eqnarray}
with the parameters $b_{i}$, $i=\overline{1-3}$, $c_{j}$, $j=\overline{1-9}$ defined in the Appendix of \cite{Oli2012} and new quantities
\begin{eqnarray}
\label{a4}
\fl A=c_{5}c_{9}-c_{6}c_{8}\\
\fl C=c_{2}c_{6}-c_{3}c_{5}\\
\fl D=c_{3}c_{8}-c_{2}c_{9}.
\end{eqnarray}
The derivatives which appear in the expressions of $M$ and $N$ are
\begin{eqnarray}
\label{a5}
\fl b_{1}'=\Lambda\eta\gamma_{2}\cos{\phi}[g\eta\cos{\phi}+G(\gamma_{2}+\Lambda)]\\
\fl b_{2}'=\gamma_{2}^{2}(\gamma_{1}+\Lambda)(\gamma_{1}+\gamma_{2})\lbrace g\Lambda\eta\lbrace\gamma_{2}(\gamma_{1}+\Lambda)(\gamma_{2}+\Lambda)\sin{\phi}\nonumber\\
\fl+[x\Delta_{2}-\eta^{2}(\gamma_{2}+\Lambda)\sin{\phi}\cos{\phi}]\cos{\phi}\nonumber\\
\fl -2G\gamma_{1}(\gamma_{2}+\Lambda)(\Delta_{1}-\Delta_{2})x\rbrace\\
\fl  b_{3}'=\gamma_{2}^{2}(\gamma_{1}+\Lambda)(\gamma_{1}+\gamma_{2})x\lbrace \Lambda\eta G(\gamma_{2}+\Lambda)\sin{\phi}\nonumber\\
\fl +g\lbrack2(\Delta_{1}-\Delta_{2})(\gamma_{2}+\Lambda)(\Lambda-\gamma_{1})+\eta^{2}\Lambda\sin{\phi}\cos{\phi}\rbrack\rbrace\\
\fl c_{1}'=2(\Delta_{1}-\Delta_{2})\gamma_{2}(\gamma_{1}+\Lambda)[(\gamma_{1}+\Lambda)(\gamma_{2}+\Lambda)-\eta^{2}\cos^{2}{\phi}]\\
\fl c_{2}'=-g^{2}[\gamma_{2}(\gamma_{1}+\Lambda)(\gamma_{1}+\gamma_{2})-\gamma_{1}\eta^{2}\cos^{2}{\phi}]+gG\eta\gamma_{2}(\gamma_{1}-\gamma_{2})\cos{\phi}\nonumber\\
\fl +2(\Delta_{1}-\Delta_{2})\gamma_{2}(\gamma_{1}+\Lambda)[\eta^{2}\sin{\phi}\cos{\phi}-\Delta_{1}(\gamma_{2}+\Lambda)]-(\gamma_{2}+\Lambda)y\nonumber\\
\fl+\eta^{2}\gamma_{2}(\gamma_{1}+\gamma_{2})^{2}+G^{2}\gamma_{2}(\gamma_{1}+\Lambda)(\gamma_{2}+\Lambda)\\
\fl c_{3}'=g^{2}(\gamma_{2}(\gamma_{1}+\Lambda))^{2}\eta\cos{\phi}+gG[\gamma_{2}(\gamma_{1}+\Lambda)(\gamma_{2}+\Lambda)-\gamma_{1}\eta^{2}\cos^{2}\phi]\nonumber\\
\fl -G^{2}\eta\gamma_{1}(\gamma_{2}+\Lambda)\cos{\phi}+2\eta(\Delta_{1}-\Delta_{2})\gamma_{2}(\gamma_{1}+\Lambda)[(\gamma_{2}+\Lambda)\sin{\phi}-\Delta_{2}\cos{\phi}]\\
\fl c_{4}'=-\gamma_{2}(\gamma_{1}+\gamma_{2})(\gamma_{1}+\Lambda)x\lbrace g^{2}[\gamma_{2}(\gamma_{1}+\Lambda)(\gamma_{1}+\gamma_{2})+\eta^{2}\cos^{2}{\phi}(\gamma_{2}-\Lambda)]\nonumber\\
\fl -gG\eta\cos{\phi}[\gamma_{2}(\gamma_{1}+\Lambda)-(2\gamma_{2}+\gamma_{1}-\Lambda)(\gamma_{2}+\Lambda)]\nonumber\\
\fl +2(\Delta_{1}-\Delta_{2})\gamma_{2}(\gamma_{1}+\Lambda)[\Delta_{1}(\gamma_{2}+\Lambda)+\eta^{2}\sin{\phi}\cos{\phi}]\nonumber\\
\fl +y(\gamma_{2}+\Lambda)-G^{2}\gamma_{2}(\gamma_{1}+\Lambda)(\gamma_{2}+\Lambda)\rbrace\\
\fl c_{5}'=\gamma_{2}(\gamma_{1}+\Lambda)^{2}g\lbrace g^{2}\lbrace (\gamma_{1}+\gamma_{2})(\gamma_{1}+\gamma_{2}-\Lambda)\eta^{2}x\sin{\phi}\cos{\phi}\nonumber\\
\fl +4(\Delta_{1}-\Delta_{2})\gamma_{2}(\gamma_{2}+\Lambda)[(\gamma_{1}+\gamma_{2})\eta^{2}\cos^{2}{\phi}-\gamma_{2}(\gamma_{1}+\Lambda)(\gamma_{1}+\gamma_{2})]\rbrace\nonumber\\
\fl +gGx\eta\sin{\phi}(\gamma_{1}+\gamma_{2})(\gamma_{1}+\gamma_{2}-\Lambda)(\gamma_{2}+\Lambda)\nonumber\\
\fl +2(\Delta_{1}-\Delta_{2})\gamma_{2}(\gamma_{1}+\Lambda)(\gamma_{1}+\gamma_{2})x[\eta^{2}\sin^{2}{\phi}-(\gamma_{1}+\Lambda)(\gamma_{2}+\Lambda)]\rbrace\\
\fl c_{6}'=-(\gamma_{2}(\gamma_{1}+\Lambda))^{2}g\lbrace g^{2}\eta(\gamma_{1}+\gamma_{2})x\lbrace(2\gamma_{2}+\gamma_{1}-\Lambda)\Delta_{2}\cos{\phi}\nonumber\\
\fl -[2\gamma_{2}(\gamma_{1}+\gamma_{2})+\Lambda(2\gamma_{2}+\gamma_{1}-\Lambda)]\sin{\phi}\rbrace\nonumber\\
\fl +gGx(\gamma_{1}+\gamma_{2})[2(\gamma_{2}+\Lambda)(\Delta_{1}-\Delta_{2})(\gamma_{1}-\Lambda)-\nonumber\\
\fl \Delta_{2}\gamma_{2}(\gamma_{1}+\Lambda)+\eta^{2}\gamma_{1}\sin{\phi}\cos{\phi}]\nonumber\\
\fl +\eta(\gamma_{1}+\gamma_{2})x\lbrace 2\gamma_{2}(\gamma_{1}+\Lambda)(\Delta_{1}-\Delta_{2})[(\gamma_{2}+\Lambda)\cos{\phi}+\Delta_{2}\sin{\phi}]\nonumber\\
\fl +G^{2}\gamma_{1}(\gamma_{2}+\Lambda)\sin{\phi}\rbrace\rbrace\\
\fl c_{7}'=\gamma_{2}(\gamma_{1}+\Lambda)(\gamma_{1}+\gamma_{2})x\lbrace g^{2}\cos{\phi}\eta\gamma_{2}(\gamma_{1}+\Lambda)\nonumber\\
\fl -gG\lbrace(2\gamma_{1}+\gamma_{2}+\Lambda)\eta^{2}\cos^{2}{\phi}-\gamma_{2}(\gamma_{1}+\Lambda)(2\gamma_{2}+\Lambda)\rbrace\nonumber\\
\fl-2(\Delta_{1}-\Delta_{2})\gamma_{2}(\gamma_{1}+\Lambda)\eta[\Delta_{2}\cos{\phi}+(\gamma_{2}+\Lambda)\sin{\phi}]\nonumber\\
\fl -G^{2}\eta(\gamma_{2}+\Lambda)(2\gamma_{1}+\gamma_{2}+\Lambda)\cos{\phi}\rbrace\\
\fl c_{8}'=\gamma_{2}(\gamma_{1}+\Lambda)\lbrace g^{2}\eta\lbrace\eta^{2}\sin{\phi}\cos^{2}{\phi}\gamma_{2}(\gamma_{1}+\gamma_{2}+2\Lambda)(\gamma_{1}+\gamma_{2})\nonumber\\
\fl -\gamma_{2}^{2}(\gamma_{1}+\Lambda)[(2\Delta_{1}\gamma_{2}+\Delta_{2}\gamma_{1})\cos{\phi}+(\gamma_{1}+\gamma_{2})(2\gamma_{2}+\gamma_{1}+2\Lambda)\sin{\phi}]\nonumber\\
\fl +\eta^{2}\gamma_{2}\Delta_{2}(\gamma_{1}+\gamma_{2})\cos^{3}{\phi}\rbrace\nonumber\\
\fl -gG\lbrace2(\Delta_{1}-\Delta_{2})\gamma_{2}(\gamma_{2}+\Lambda)[\gamma_{2}(\gamma_{1}+\Lambda)(2\gamma_{1}+\gamma_{2}+\Lambda)-(\gamma_{1}+\gamma_{2})\eta^{2}\cos^{2}{\phi}]\nonumber\\
\fl -x\eta^{2}(\gamma_{1}+\gamma_{2})(2\gamma_{1}+\gamma_{2}+\Lambda)\sin{\phi}\cos{\phi}\nonumber\\
\fl +\gamma_{2}(\gamma_{1}+\Lambda)\lbrace(\gamma_{1}+\gamma_{2})\eta^{2}\cos{\phi}(\Delta_{2}\cos{\phi}-\gamma_{2}\sin{\phi})\nonumber\\
\fl +\gamma_{2}(\gamma_{1}+\Lambda)[\Delta_{2}(2\gamma_{2}-\gamma_{1}+2\Lambda)-2\Delta_{1}(2\gamma_{2}+\Lambda)]\rbrace\rbrace\nonumber\\
\fl -\eta\lbrace2(\Delta_{1}-\Delta_{2})\gamma_{2}(\gamma_{1}+\Lambda)\lbrace(\gamma_{2}+\Lambda)\lbrace\gamma_{2}(\gamma_{1}+\Lambda)[(\Delta_{1}-\Delta_{2})\sin{\phi}\nonumber\\
\fl +(\gamma_{1}+\gamma_{2})\cos{\phi}]-\eta^{2}(\gamma_{1}+\gamma_{2})\cos{\phi}\rbrace\nonumber\\
\fl +\sin{\phi}\lbrace\eta^{2}\cos{\phi}(\gamma_{1}+\gamma_{2})[(\gamma_{2}+\Lambda)\sin{\phi}+\Delta_{2}\cos{\phi}]\nonumber\\
\fl -\gamma_{2}(\gamma_{1}+\Lambda)[\Delta_{2}(\gamma_{1}-\Lambda)+\Delta_{1}(2\gamma_{2}+\Lambda)]\rbrace\rbrace\nonumber\\
\fl -y\gamma_{2}^{2}(\gamma_{1}+\Lambda)-G^{2}(\gamma_{2}+\Lambda)(2\gamma_{1}+\gamma_{2}+\Lambda)(\gamma_{1}+\gamma_{2})x\sin{\phi}\rbrace\rbrace\\
\fl c_{9}'=x(\gamma_{1}+\gamma_{2})\gamma_{2}(\gamma_{1}+\Lambda)\lbrace2g^{2}\gamma_{2}(\gamma_{1}+\Lambda)\Delta_{2}\nonumber\\
\fl +gG\eta[(\gamma_{1}+\gamma_{2})(\gamma_{1}+\gamma_{2}+\Lambda)\sin{\phi}-(3\gamma_{1}+\gamma_{2}+\Lambda)\Delta_{2}\cos{\phi}]\nonumber\\
\fl -2(\Delta_{1}-\Delta_{2})\lbrace\Delta_{2}^{2}\gamma_{2}(\gamma_{1}+\Lambda)+(\gamma_{2}+\Lambda)[G^{2}(2\gamma_{1}+\Lambda)+\gamma_{2}(\gamma_{1}+\Lambda)(\gamma_{2}+\Lambda)]\rbrace
\end{eqnarray}
with $x$ and $y$ being
\begin{eqnarray}
\label{a6}
\fl x=\gamma_{2}(\gamma_{1}+\Lambda)-\eta^{2}cos^{2}\phi\\
\fl y=\gamma_{2}(\gamma_{1}+\Lambda)[(\gamma_{1}+\gamma_{2})^{2}+(\Delta_{1}-\Delta_{2})^{2}]-\eta^{2}(\gamma_{1}+\gamma_{2})^{2}.
\end{eqnarray}


\section*{References}
\begin{thebibliography}{14} 
\bibitem{hau} Hau L V, Harris S E, Dutton Z and Behroozi C H 1999 {\it Nature (London)} \textbf{397} 594
\bibitem{fleischhauer} Fleischhauer M and Lukin M D 2000 {\it Phys. Rev. Lett.} \textbf{84} 5094
\bibitem{phillips} Phillips D F, Fleischhauer A, Mair A, Walsworth R L and Lukin M D 2001 {\it Phys. Rev. Lett.} \textbf{86} 783
\bibitem{liu} Liu C, Dutton Z, Behroozi C H and Hau L V 2001 {\it Nature (London)} \textbf{409} 490
\bibitem{turukhin} Turukhin A V, Sudarshanam V S, Shahriar M S, Musser J A, Ham B S and Hemmer P R 2001 {\it Phys. Rev. Lett.} \textbf{88} 023602
\bibitem{ku} Ku P C, Chang-Hasnian C J and Chuang S L 2002 {\it Electron. Lett.} \textbf{38} 1581
\bibitem{mikhailov} Mikhailov E E, Sautenkov A V, Rostovtsev Y V and Welch R G 2004 {\it J. Opt. Soc. Am.} B \textbf{21} 425
\bibitem{tseng} Tseng H Y, Huang J and Adibi A 2006 {\it Appl. Phys.} B \textbf{85} 493
\bibitem{dahan} Dahan D and Eisenstein G 2005 {\it Opt. Express} \textbf{13} 6234
\bibitem{longdell} Longdell J J, Fraval E, Sellars M J and Manson N B 2005 {\it Phys. Rev. Lett.} \textbf{95} 063601 
\bibitem{wu} Wu Y and Deng L 2004 {\it Phys Rev Lett} \textbf{93} 143904
\bibitem{wang} Wang L J, Kuzmich A and Dogariu A 2000 {\it Nature (London)} \textbf{406} 277
\bibitem{chiao} Chiao R Y 1993 {\it Phys. Rev.} A \textbf{48} R34
\bibitem{peatross} Peatross J, Glasgow S A and Ware M 1999 {\it Phys. Rev. Lett.} \textbf{84} 2370 
\bibitem{kim} Kim K, Moon H S, Lee C, Kim S K and Kim J B 2003 {\it Phys. Rev.} A \textbf{68} 013810
\bibitem{bae} Bae I-H and Moon H S 2011 {\it Phys. Rev.} A \textbf{83} 053806
\bibitem{bigelow} Bigelow M S, Lepeshkin and Boyd R W 2003 {\it Science} \textbf{301} 200
\bibitem{agarwal} Agarwal G S, Dey T N and Menon S 2001 {\it Phys. Rev.} A \textbf{64} 053809
\bibitem{han} Han D, Guo H, Bai Y and Sun H 2005 {\it Phys. Lett.} A \textbf{334} 243
\bibitem{joshi} Joshi A, Hassan S S and Xiao M 2005  {\it Phys. Rev.} A \textbf{72} 055803
\bibitem{Jav}Javanainen J 1992 {\it Europhys. Lett.} {\bf 17} 407
\bibitem{mahmoudi} Mahmoudi M, Sahrai M and Tajalli H 2006 {\it J. Phys. B: At. Mol. Opt. Phys.} \textbf{39} 1825; Mahmoudi M, Sahrai M and Tajalli H 2006 {\it Phys. Lett.} A \textbf{357} 66
\bibitem{dastidar} Dutta S and Dastidar R 2007 {\it J. Phys. B: At. Mol. Opt. Phys.} \textbf{40} 4287
\bibitem{guo} Bai Y, Guo H, Han D and Sun H 2005 {\it Phys. Lett.} A \textbf{357} 66
\bibitem{arbiv} Bortman-Arbiv D, Wilson-Gordon A D and Friedmann H 2001 {\it Phys. Rev.} A \textbf{63} 043818
\bibitem{hanD} Han D, Zeng Y, Bai Y, Chen W and Lu H 2006 {\it J. Mod. Opt.} \textbf{54} 493 
\bibitem{Oli2012} Budriga O 2012 {\it Eur. Phys. J.} D \textbf{66} 137
\bibitem{ficek} Ficek Z and Swain S 2004 {\it Quantum Interference and Coherence} (Berlin: Springer) p 258-259 
\bibitem{OliOptCommun2014} Budriga O 2014 {\it Opt. Commun.} DOI: 10.1016/j.optcom.2014.04.064 
\bibitem{cardimona}Cardimona D A, Raymer M G, Stroud Jr C R 1982 {\it J. Phys. B: At. Mol. Phys.} \textbf{15} 55
\bibitem{Agarwal} Menon S and Agarwal G S 1998 {\it Phys. Rev.} A {\bf 57} 4014
\bibitem{Paspalakis} Paspalakis E, Gong S Q and Knight P L 1998 {\it J. Mod. Opt.} {\bf 45} 2433
\bibitem{Oli2013} Budriga O 2013 {\it Phys. Scr.} \textbf{T153} 014007
\bibitem{physrevlet77} Xia H R, Ye C Y and Zhu S Y 1996 {\it Phys. Rev. Lett.} \textbf{77} 1032
\bibitem{hakuta}Hakuta K, Marmet L and Stoicheff B P 1991 {\it Phys. Rev. Lett.} \textbf{66} 596
\bibitem{faist}Faist J, Capasso F, Sirtori C, West K V and Pfieffer L N 1997 {\it Nature (London)} \textbf{390} 589
\bibitem{berman}Berman P R 1998 {\it Phys. Rev.} A \textbf{58} 4886
\bibitem{patnaik}Patnaik A K and Agarwal G S 1999 {\it Phys. Rev.} A \textbf{59} 3015
\bibitem{optcomun179}Zhou P and Swain S 2000 {\it Opt. Commun.} \textbf{179} 267
\bibitem{anindita} Bhattacharjee A, Sanyal S and Dastidar K R 2005 {\it J. Mol. Spectrosc.} \textbf{232} 264 
\bibitem{Cohen} Cohen-Tannoudji C, Dupont-Roc J and Grynberg G 2004 {\it Atom-Photon Interactions: Basic Processes and Applications} (Weinheim: Wiley-VCH Verlag) p 415-418


\end{thebibliography}
\end{document}